\documentclass[aps,pra,floatfix,twocolumn,superscriptaddress]{revtex4-1}
\usepackage[final]{graphicx}
\usepackage{times,amsmath,amssymb}
\usepackage{epsfig,color}
\usepackage{hyperref}
\usepackage[caption = false]{subfig}
\usepackage{thumbpdf,enumerate}
\usepackage{booktabs}
\usepackage{bbm} 

\newcommand{\Tr}{{\rm Tr}}

\def\Tr{\hbox{Tr}} 
\usepackage{pifont}

\newcommand{\ket}[1]{\vert#1\rangle}
\newcommand{\bra}[1]{\langle#1\vert}

\begin{document}

\title{Quantum Coherence in Noisy Cellular Automata}

\author{Federico Centrone}
\affiliation{Dipartimento di Fisica, Sapienza Universit\`a di Roma , Piazzale Aldo Moro 5, 00185 Rome, Italy.}

\author{Marco Barbieri}
\affiliation{Dipartimento di Scienze, Universit\`{a} degli Studi Roma Tre, Via della Vasca Navale 84, 00146, Rome, Italy}

\author{Alessio Serafini}
\affiliation{Department of Physics and Astronomy, University College London, Gower Street, London WC1E 6BT, UK}

\begin{abstract}
Quantum cellular automata are important tools in understanding quantum dynamics, thanks to their simple and effective list of rules. Here we investigate 
explicitly how coherence is built and lost in the evolution of one-dimensional automata subject to noise. Our analysis illustrates the interplay between unitary and noisy dynamics, and draws considerations on their behaviour in the pseudo-frequency domain.
\end{abstract}

\maketitle

The simulation of physical phenomena can be a demanding task. The technical efforts that need being devoted to making physical phenomena {\it in silico} easible have originated into the field of computational physics. Cellular automata have been introduced in this context in the early days of computer science in order to investigate self-replicating systems, as well as other peculiar discrete dynamical occurrences~\cite{bib:vonNeu1}. Classical cellular automata are built as arrays of elementary cells, being in either an active or a passive state: at each time step the state of all the cells is updated depending on the state of the surrounding ones. Cellular automata then provide one with a test system whose evolution can be studied by invoking a minimal set of rules, thus considerably reducing technical requirements on the simulation. Despite their simplicity, complex unpredictable behaviours can be observed, as in the celebrated example of Conway's game of life~\cite{bib:GoL}, or as illustrated in the extensive work of Wolfram~\cite{bib:wolfram, bib:Aaronson}. Nowadays, cellular automata have found connections to random number generation~\cite{bib:wolfram3}, cryptography~\cite{bib:wolfram4}, universal computing~\cite{bib:cook}, and there even exist proposals for a unifying string theory based on this model~\cite{bib:Hooft}. 

Quantum features can be introduced as a way to gain insight on a larger class of fundamental effects: quantum cellular automata (QCA) allow for superposition states for the elementary cells~\cite{bib:schumacher,watrous}. Like their classical counterparts, QCA have been studied for their implications for quantum computing~\cite{Raussendorf,Vollbrecht,Nagaj,Arrighi12,Dariano}, but they have also found application to fundamental studies on quantum fields~\cite{Lloyd, Arrighi, Mosco, Bisio}. The evolution of the whole array is generally taken to be unitary, however there exist restrictions on the acceptable propagators, since they need to satisfy requirements on discreteness in both time and space, homogeneity, locality, and causality.

\begin{figure}[b!]
\includegraphics[width= \columnwidth, trim={0 2.5cm 0 1cm}, clip]{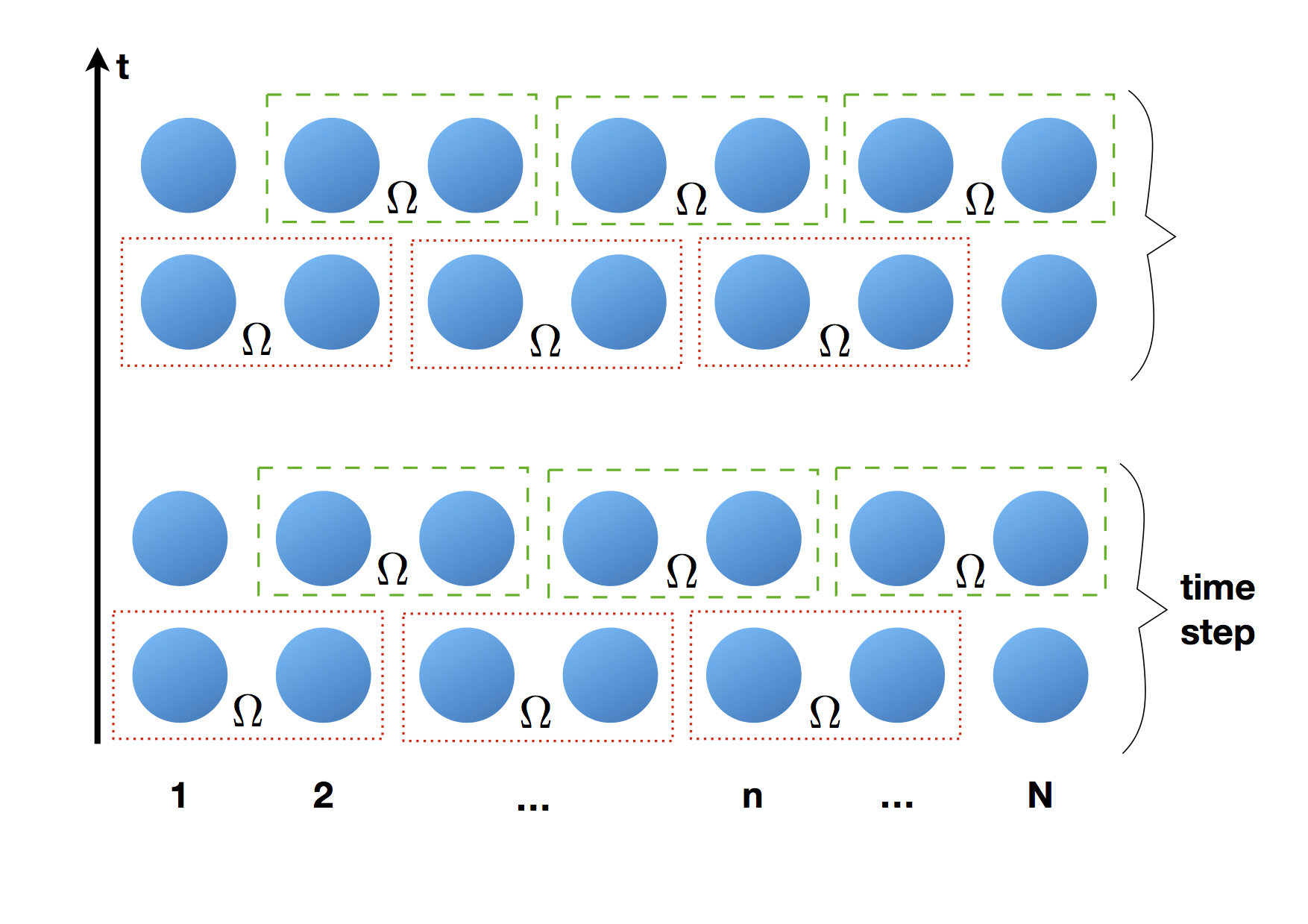}
\caption{One-dimensional  quantum cellular automata. Each node is a two-level system, and excitations can be transferred between adjacent nodes by the action of the quantum map $\Omega$, that can be tuned from unitary to classical stochastic by the amount of noise present. A single time step of the full QCA evolution consists in replicating the action of $\Omega$ on the red pairings first (dotted boxed), and then on the green pairings (dashed boxes).}
\label{fig:array}
\end{figure}

QCA are closely related to quantum walks~\cite{Aharonov, Perets, Peruzzo, Schreiber, Kitagawa, Crespi, Cardano, Defienne}, since both architectures can explore the propagation of excitations and quantum information through networks; in this respect, it should be possible to have a unifying picture of classical and quantum evolutions by introducing tuneable noise. This has been accomplished in~\cite{Avalle1}, where a suitable generalisation of unitaries to generic maps is illustrated. It has been shown that excitation transfer in the QCA array can be continuously driven from the quantum to the classical stochastic regime. The presence of quantum coherence, in general, ensures a higher transfer efficiency on shorter times, sometimes with the assistance of noise~\cite{Avalle1}. Related studies have also examined quantum information processing via noisy QCA~\cite{Avalle2,McNally}. 
It should be noted that non-unitary quantum automata had already been introduced in \cite{brennen03}.

These studies are part of an effort towards understanding how quantum properties may affect and possibly enhance transport in networks, with possible implications for biophysics~\cite{quantumtransport}. On the other hand, such an interest in quantum coherence, compounded with the general framework of resource theories proper to quantum information science, has led to the introduction of suitable quantifiers for quantum coherence as a genuine resource~\cite{Baumgratz,Levi, Streltsov,Winter,RMP}.

Since the noisy QCA construction of Ref.~\cite{Avalle1} was meant to explore the passage from classical to quantum dynamics, 
it is interesting to analyse the evolution of quantum coherence is such a model in view of the recent quantifiers mentioned above. 
Therefore, we use the measures of quantum coherence introduced in \cite{Baumgratz} to investigate how this is produced in the evolution of a linear-array QCA. Our studies found that in the noiseless case coherence oscillates in time with a spectrum of frequencies which depends on the size. In the presence of relevant classes of noise, namely dephasing and amplitude damping, we assist to a built of the coherence at early times, followed by an exponential decay for long times. Remarkably, the decay is slower for increasing size.
 
Our model of QCA is a one-dimensional array of $N$ two-level nodes, being in either an exited or a ground state; these represent the elementary cells of the automaton. We consider here the case when a single excitation is present at the initial time $t=0$. This restricts our attention to the single-excitation sector: the possible classical states of the network, the ones in which the excitation is present at the $n$-th site, are labelled as $\ket{n}$. For the dynamics to capture the essential features of the automata, it needs to satisfy the above mentioned requirements of discreteness in time and space, homogeneity, locality, and causality: these are met if we partition the system in pairs of neighbouring nodes, 1 and 2, 3 and 4, and so on (Fig.\ref{fig:array}), and update the state of the QCA via unitary operations in the form:
\begin{equation}
U =\left( 
\begin{matrix}
\cos\theta &  \sin\theta \,e^{i\phi_2} \\
-\sin\theta \,e^{i\phi_1} &  \cos\theta \,e^{i(\phi_1+\phi_2)}
\end{matrix}
\right)
\label{eqn:U}
\end{equation}

\begin{figure}[t!]
\includegraphics[width= \columnwidth ]{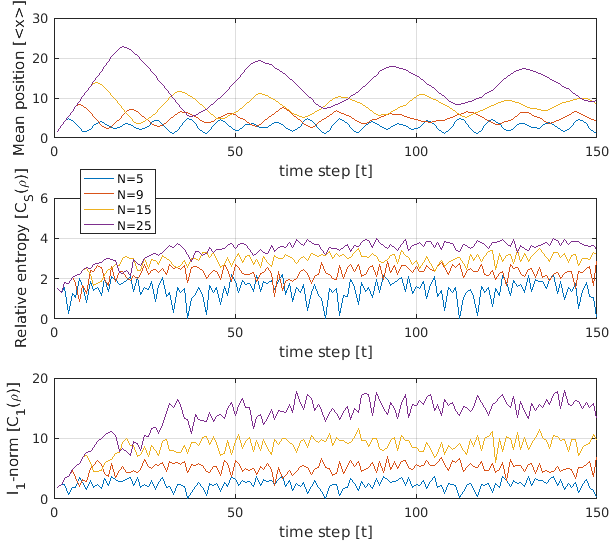}
\caption{Evolution of the quantum coherence of the noiseless QCA as a function of the time step for different lengths of the array. The unitary evolution considered corresponds to $\theta{=}\pi/4$ in Eq.\eqref{eqn:U} the one giving the highest value. Upper panel: average position of the excitation $\langle x \rangle$. Middle panel: evolution of the entropy-based measure of coherence $C_S(\ket{\psi}\bra{\psi})$. Lower panel: evolution of the $\l_1$-norm measure $C_1(\ket{\psi}\bra{\psi})$.}
\label{fig:coherenceNN}
\end{figure}

\begin{figure}[t]
\includegraphics[width=0.9\columnwidth]{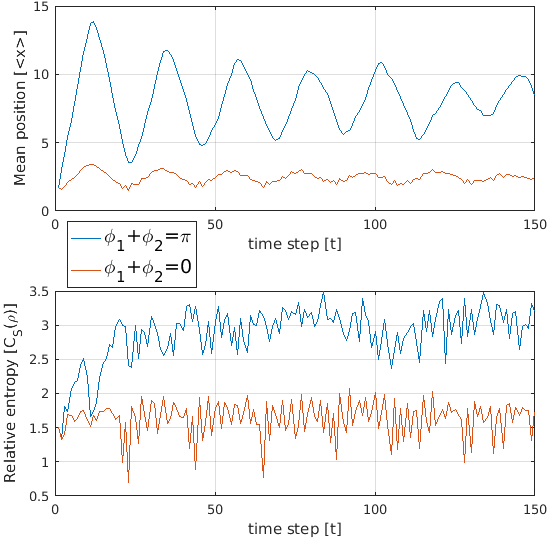}
\caption{The role of the phase $\phi_1+\phi_2$ in the production of quantum coherence, without noise and for $\theta=\pi/4$. The length of the array is $N=15$. The occurrence of localised states in the array for $\phi_1+\phi_2=\pi$ prevents to achieve maximal coherence (blue line), while for $\phi_1+\phi_2=0$ the localisation mechanism is suppressed, and the QCA can explore its full space.}
\label{fig:coherencePhi}
\end{figure}

\noindent on the qubit represented by $\ket{n}$ and $\ket{n{+}1}$ for all the pairings above: this amounts to transferring the excitation with probability $\sin^2\theta$ to the adjacent node, and leaving on the original node with probability $\cos^2\theta$. This action is different from a simple stochastic transfer since phases are established between these two possibilities. The time-step is completed by shifting the partitioning by one cell to the right (Fig. \ref{fig:array}), and apply $U$ on the pairs of nodes 2 and 3, 4 and 5, {\it et cetera}, in such a way that the whole network is consider. 

This purely unitary dynamics can be generalised  to a dissipative one by introducing dephasing and amplitude damping to the qubit evolution; these have been shown to be sufficient to reproduce stochastic excitation transfer in the chain~\cite{Avalle1}. Dephasing is described by the map $\Phi_\xi$, charaterised by a parameter $\xi$ describing the strength of the dephasing; the associated Krauss operators are
\begin{equation}
D_0=\sqrt{1-\xi} \mathbbm{1},\quad D_1=\sqrt{\xi}( \mathbbm{1}+\sigma_z)/2 ,\quad   D_2= \sqrt{\xi}(\sigma_z- \mathbbm{1}),
\label{eqn:D}
\end{equation}
where $\sigma_j$ for $ j = x, y, z $ stand for the Pauli matrices. Along with dephasing, we will also consider amplitude damping $\Xi_\eta$, with strength $\eta$, and associated operators 
\begin{equation}
L_{0, \eta}=\frac{\mathbbm{1}+\sigma_z}{2} +\sqrt{1-\eta}\frac{\mathbbm{1}-\sigma_z}{2} \quad  L_{1, \eta} = \sqrt{\eta}\frac{\sigma_x+i\sigma_y}{2}.
\label{eqn:L}
\end{equation}
While normally $0\leq\eta\leq 1$, we can extend this to negative values ($-1\leq\eta< 0$) as a shorthand notation for the inverted channel, {\it i.e.} the one with elements $\sigma_x L_{0, |\eta|}\sigma_x$, and $\sigma_x L_{1, |\eta|}\sigma_x$. Overall, the action of the complete map on a generic qubit state $\rho$ is written as:
\begin{equation}
\Omega_{\xi,\eta}(\rho)=\Xi_\eta\left(\Phi_\eta\left(U\rho U^\dag \right)\right),
\label{eqn:O}
\end{equation}
so that the evolution of the full network is a composition of these maps applied to node pairs, as described above (Fig.\ref{fig:array}). The complete map $\Omega_{\xi,\eta}$ can be reparametrised by defining $\eta = p-q$, and $\cos(2\theta)=(1-p-q)/(1-|\eta|)$. 

\begin{figure}[t!]
\includegraphics[width= \columnwidth]{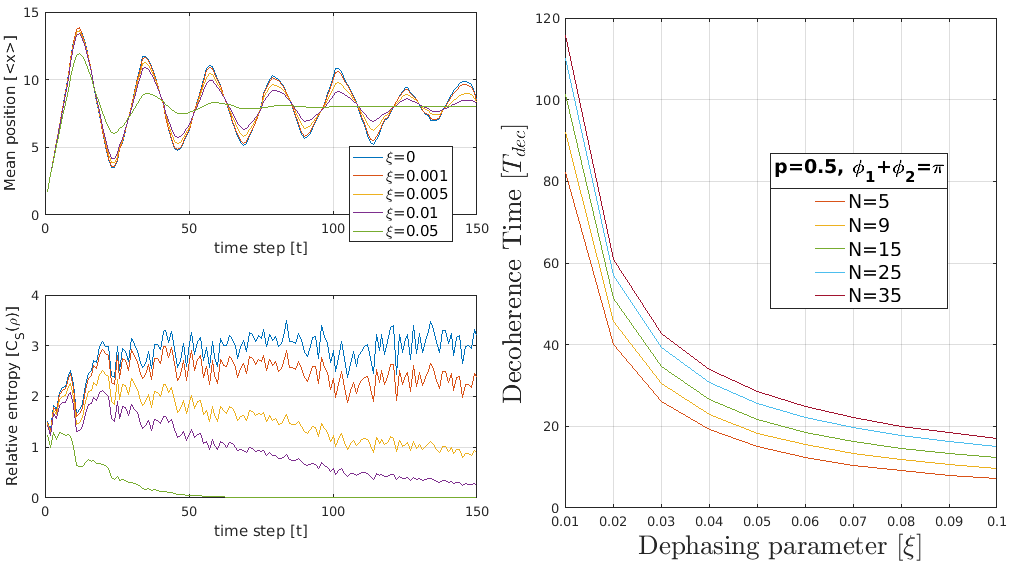}
\caption{Coherence in the presence of pure dephasing. The panels on the left consider a chain of length $N=15$, with the phase in the unitary satisfying $\phi_1+\phi_2=\pi$.}
\label{fig:noise}
\end{figure}

Coherence in a quantum system in the state $\rho$ can be quantified by means of the relative entropy with respect the state $\rho^D$, which has the same populations as $\rho$, but all the off-diagonal terms are zero~\cite{Baumgratz}:
\begin{equation}
\label{eq:cs}
C_S(\rho)=S( \rho||\rho^D )=\Tr(\rho\ln\rho)-\Tr(\rho^D\ln\rho^D).
\end{equation}
This implies that we have determined a privileged basis for decomposing the matrix $\rho$ on physical motivations; in our system this is naturally given by the $\ket{n}$ vectors. The measure $C_S(\rho)$ is then interpreted as the amount of information that is needed to learn $\rho$ if one has knowledge of its diagonal terms. Alternatively, one can build a measure based on the $\l_1$ distance between $\rho$ and $\rho^D$ ~\cite{Baumgratz}:
\begin{equation}
\label{eq:c1}
C_1(\rho)=\| \rho-\rho^D \|_1=\sum_{i\neq j}|\rho_{i,j}|,
\end{equation}
which is simply the magnitude of the off-diagonal terms. The notion of coherence can then be extended to quantum operations, by defining as incoherent those maps that can not generate coherent states from incoherent states~\cite{Aberg,Chitambar,Chitambar1,Marvian}. Both measures satisfy the requirements of vanishing for incoherent states, not increasing under incoherent operations, and representing a proper distance between $\rho$ and the closest incoherent state~\cite{Baumgratz}.

We start our analysis with the evolution in the absence of any decoherence effect, thus setting $\eta=0$, and $\xi=0$. The array is initialised in the state $\ket{1}$, {\it i.e.} the excitation is present only on the first node; at each time step, the full chain is therefore in a pure state $\ket{\psi}=\sum_n c_n \ket{n}$. The evolution of the coherence as a function of the time steps is illustrated in Fig.\ref{fig:coherenceNN}:  the coherence reaches a limit value, in the presence of fast oscillations. As one might expect, the two measures \eqref{eq:cs} and \eqref{eq:c1} display similar behaviours, both of them increasing with the number of nodes $N$, 
the former with $\ln N$, the latter with $N$. This is due to the fact that the dimension of the single-excitation subspace to which the evolution is constrained is $N$, while the maxima taken by the two measures over such a space are, respectively, $\ln(N)$ and $N-1$. 
Notice that this applies to the case $\theta=\pi/4$ which, corresponding to a balanced superposition at the single-qubit level, spreads out the coherence maximally over the chain and thus attains maximum coherence on the shortest time-scale. 
The oscillations in the coherence are connected to the average position $\langle x \rangle=\sum_n |c_n|^2n$ of the excitation in the chain: the minima of $C_S(\ket{\psi}\bra{\psi})$ and $C_1(\ket{\psi}\bra{\psi})$ occurring in correspondence of the extrema of $\langle x \rangle$. Since the two measures give similar qualitative information, in the following we can then focus on $C_S(\rho)$.

It has been observed that for given coupling strengths in the unitary $U$ in \eqref{eqn:U}, {\it i.e.} for a given $\theta$, the QCA presents different behaviours depending on the sum of the phases $\phi_1{+}\phi_2$: if the latter is zero, localisation of the excitation might occur. This is reflected in the coherence, as shown in Fig.~\ref{fig:coherencePhi}: localisation restricts the evolution of the system to a subset of its accessible states, thus limiting the maximal value of $C_S(\ket{\psi}\bra{\psi})$, and $C_1(\ket{\psi}\bra{\psi})$. In both cases, similar short-time oscillations occur.

\begin{figure}[b]
\includegraphics[width= \columnwidth]{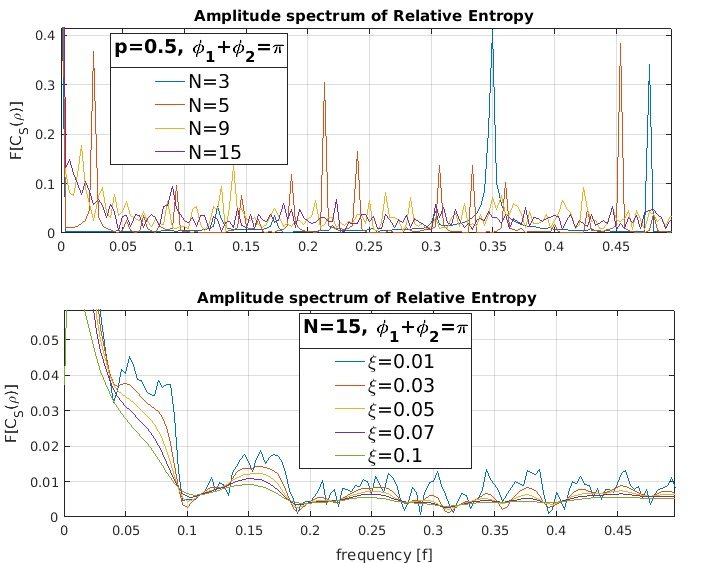}
\caption{Fast Fourier Transform of the coherence $C-S(\rho(t))$ in the noiseless (upper) and noisy (lower panel) case. Notice how increasing the size $N$ in the absence of noise leads to a complex structure of the spectrum, due to building of coherence terms at different time scales. In the presence of noise, long-term oscillations are inhibited by the presence of noise, and, in general, the shape of the spectrum is simplified.}
\label{fig:fourier}
\end{figure}

\begin{figure*}[t]
\includegraphics[width= \textwidth]{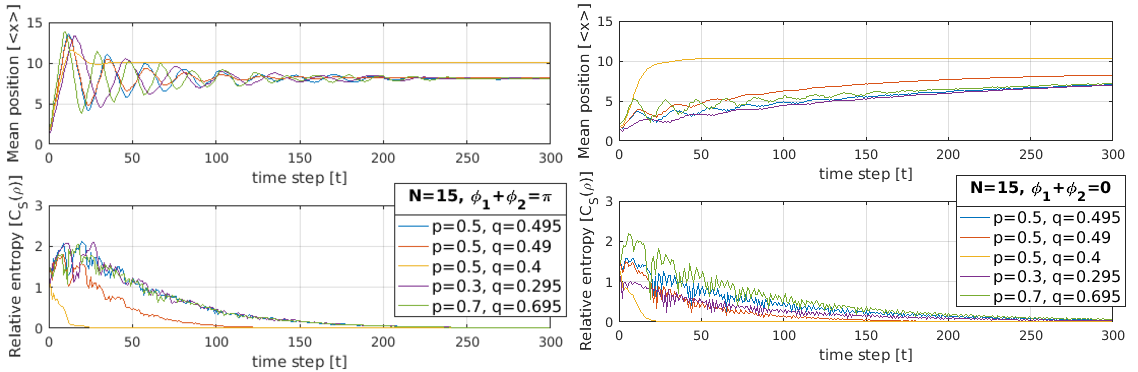}
\caption{Decoherence in the presence of amplitude damping. The plots report the mean position of the excitation, and the degree of coherence $C_S(\rho)$ without (left column) and with (right column) localisation effects. The damping is expressed in terms of $\eta=p-q$.}
\label{fig:damping}
\end{figure*}

We now turn our attention to how noise affects this behaviours; we will start reporting the effect of the dephasing strength $\xi$, with no amplitude damping $\eta=0$. The QCA is inisitalised in the same pure state $\ket{1}$ as before, but now dephasing occurs during its evolution. At any time, the system will be in a mixture $\rho(t)$. When inspecting the coherence $C_S(\rho(t))$, a competition between two effects is observed, Fig.~\ref{fig:noise}a: while at short times coherence is built by the action of the unitary $U$, for longer times this starts being reduced by the presence of the noise. The average position of the excitation reflects this interplay: its oscillations are damped as coherence vanishes. The QCA is then driven to a long-term completely incoherent state, representing the thermal death of the system~\cite{trailer}. A characteristic time $T_{\rm dec}$ for the decoherence can be estimated by fitting $C_S(\rho)$ with an exponential curve ${\sim}e^{-t/T_{\rm dec}}$. This is reported as a function of $\xi$ for different chain lengths $N$ (Fig.~\ref{fig:noise}b): the same level of noise results in a slower loss of coherence as the size of the QCA increases. This can be understood by considering the action of the dephasing  Kraus operators on the direct sum Hilbert space that forms the single-excitation sector: on a single application 
of the map, only the off-diagonal elements pertaining 
to pairs of nearest neighbour sites are damped, whilst the coherence between farther removed sites is not affected at all. Hence, coherence is damped only via the dephasing of the elements immediately above (and below) the main diagonal. Longer chains are therefore comparatively less affected by such a noise for the same strength. Notice that this still damps the overall coherence, since the off-diagonal elements farther away from the main diagonal must still build up from the damped ones through repeated applications of the unitary part of the local automaton.

Insight can be gained by inspecting the Fourier transform of $C_S(\rho(t))$, thus considering the pseudo-frequency domain with respect to the evolution time (Fig.~\ref{fig:fourier}).  In the noiseless case, the spectrum presents distinct peaks only for short array lengths. As the number of nodes increases, the spectrum starts presenting a large tail with rapid modulations. This reflects how correlations are established between nodes of any distance: for growing $N$ this results in multiple time scales being introduced in the problem. When adding the action of dephasing, the spectrum is smoothed, and low-frequency components are strongly suppressed.

Similar considerations can also be derived when inspecting the behaviour under amplitude damping (we fix $\xi=0$): Fig.~\ref{fig:damping} illustrates that the automata present a very similar trend in its coherence as under dephasing, with the same peculiar distinction between constructive and destructive interference effect that lead, in the latter case, to localisation, which is not disrupted by the presence of damping.

As a concluding remark, we observe that the loss of coherence occurs rapidly even for modest levels of noise: in Figs. \ref{fig:noise} and \ref{fig:damping} it is evident how values of the noise parameters as low as 0.1 are sufficient to confine a coherent behaviour to the first few tens of time steps. Therefore, coherent dynamics itself seems to be only relevant for fast processes.

{\it Acknowledgements}. We thank P. Mataloni, M. Paris and M. Avalle for discussion and encouragement.

\end{document}